\documentstyle[multicol,epsfig,epsf,aps,prl]{revtex}
\begin{document}
\draft
\tightenlines
\title{Level-statistics of Disordered Systems: a single parametric formulation}
\author{Pragya Shukla$^{*}$}
\address{(i) Max-Planck Institute for Physics of Complex Systems, 
Nothnitzer Str.38, Dresden-01187, Germany, \\
(ii) Department of Physics,
Indian Institute of Technology, Kharagpur-721302, India.}
\onecolumn
\date{\today}
\maketitle
\widetext
\begin{abstract}

	We find that the statistics of levels undergoing 
metal-insulator transition in systems with multi-parametric Gaussian 
disorders behaves in a way similar to that of the single parametric 
Brownian ensembles 
\cite{dy}. The latter appear during a     
  Poisson $\rightarrow$ Wigner-Dyson transition, driven by a 
random perturbation.  
 The analogy  provides  
the analytical evidence for the single parameter scaling behaviour in 
disordered systems as well as a tool to    
 obtain the level-correlations at the 
critical point for a wide range of disorders.

\end{abstract}

.
\begin{multicols}{2}

	The nature of the eigenfunctions can significantly affect the 
statistical behaviour of energy levels of a disordered system and thereby 
related physical properties e.g transport. The presence of disorder may cause 
localized waves in the system, implying lack of interaction between 
certain parts. 
This is reflected in the structure of the Hamiltonian matrix which is 
 sparse or banded in the site representation depending on the 
dimensionality of the system.   	
	The variation of the disorder-strength can lead to a metal-insulator 
transition (MIT), with eigenfuctions changing from a 
fully extended state (metal) to a strongly localized one (insulator) with 
partial localization in the critical region. The 
associated Hamiltonian also undergoes a transition from a full matrix 
to a sparse or banded form and finally to a diagonal matrix.
The statistical studies of levels for various types of disorders 
require, therefore, analysis of different ensembles. Here the nature of the 
localization and its strength is reflected in the measure and the sparcity 
of the ensemble, respectively.  Our objective in this paper 
is to obtain a mathematical formulation for the level-correlations, common 
to a large class of disorders (Gaussian type); the information 
about the nature of disorder enters in the formulation 
through a parameter, basically 
a function of various parameters influencing the localization.   

	Recently it was shown that the eigenvalue distributions
of various ensembles, with a Gaussian measure,  appear as non-equilibrium 
stages of a Brownian type diffusion process \cite{ps}.
Here the eigenvalues evolve with respect to a
parameter related to the complexity of the system represented by the ensemble.
The solution of the diffusion equation
for a given value of the parameter gives, therefore, the
distribution of the eigenvalues, and thereby their correlations, for the 
corresponding system. The present study uses the technique to analyze the
spectral properties of the levels undergoing MIT in various disordered systems.

	The Anderson model for a disordered system is described by
 a d-dimensional disordered lattice, of size $L$, with a
  Hamiltonian $H= \sum_n \epsilon_n a_n^+ a_n -
\sum_{n\not=m} b_{mn} (a_n^+ a_m +a_n a_m^+)$ in tight-binding
approximation.
The site energies $\epsilon_n$, measured in units of the
overlap integral between adjacent sites, correspond to the random potential.
The hopping is  
assumed to connect only the $z$ nearest-neighbors (referred by $m$) 
 of each site. 
In the site representation, $H$
turns out to be a sparse matrix of size $N=L^d$ with diagonal
matrix elements $H_{kk}=\epsilon_k$ and off-diagonals $H_{kl}$ describing 
the hopping.
 The level-statistics can
therefore be studied by analyzing the properties of an ensemble of (i) sparse
real symmetric matrices, in presence of a time-reversal symmetry  and
(ii) sparse complex Hermitian matrices in absence of a time-reversal.

The MIT can be brought about by a competitive variation of the
disorder and hopping rate which can be mimicked by a change of the
distribution parameters of the ensemble measure $\rho(H)$.
In this paper, we consider the case in which 
the site-energies $\epsilon_i$ are independent Gaussian distributions. 
The hopping can be chosen to 
be non-random or random (Gaussian). The $\rho(H)$, for any intermediate 
state of MIT can therefore be described by

\begin{eqnarray}
 \rho (H,y,b)=C{\rm exp}[{-\sum_{s=1}^\beta \sum_{k\le l} (1/2 h_{kl;s}) 
 (H_{kl;s}-b_{kl;s})^2 }]
\end{eqnarray}
 with subscript $"s"$ of a variable referring to its components, 
$\beta$ as their 
total number ($\beta=1$ for real variable, $\beta=2$ for the complex one), 
$C$ as the normalization constant, $h$  as the set of the variances 
$h_{kl;s}=<H^2_{kl;s}>$
 and $b$ as the set of all $b_{kl;s}$.
As obvious, in the limit $h_{kl;1}, h_{kl;2} \rightarrow 0$,
eq.(1) corresponds to the
non-random nature of $H_{kl}$ ($H_{kl}= b_{kl;1} + i b_{kl;2}$).
Here the strongly insulated state corresponds to all $h_{kl}\rightarrow 0$ 
($k\not= l$) and $b_{kl}\rightarrow 0$  (all $k,l$), implying no 
overlap between site states. The metal, with almost similar overlap 
between various site states, can be modeled by eq.(1) with all 
$h_{kl}\rightarrow\gamma^{-1}$, that is, a 
Wigner Dyson (WD) ensemble \cite{dy}.

The  eigenvalue distribution for a metal is given by the WD distribution, 
and, for an insulator by a Poisson distribution \cite{ss}. 
The  distribution for 
various transition stages can be 
obtained by integrating $\rho$ over the associated eigenvector space.
Let $P(\mu,h,b)$ be the probability of finding eigenvalues
$\lambda_i $  of $H$ between $\mu_i$ and $\mu_i+{\rm d}\mu_i$ at
a given $h$ and $b$, it can be expressed as
$P(\mu,h,b)= \int \prod_{i=1}^{N}\delta(\mu_i-\lambda_i) \rho (H,h,b){\rm d}H$.
As discussed in ref.\cite{ps}, it is possible to define a "complexity"
parameter $Y$, a function of various distribution parameters $h_{kl;s}$ and 
$b_{kl;s}$ \cite{ps}, 
\begin{eqnarray}
Y= -{1\over 2 M \gamma} 
 {\rm ln}\left[ \prod_{k \le l}^{'}\prod_{s=1}^{\beta}
 |x_{kl;s}| \quad |b_{kl;s}|^2 \right] + C 
\end{eqnarray}
(here the $\prod'$ implies a product over non-zero $b_{kl;s}$ and 
$x_{kl;s} \equiv 1-\gamma g_{kl} h_{kl;s}$ only, $g_{kl}=2-\delta_{kl}$,
$C$ as a constant determined by initial ensemble and  
$M$ as the number of all non-zero 
parameters $x_{kl;s}$ and $b_{kl;s}$) such that the evolution
of $P$ with respect to $Y$ results in the diffusion of eigenvalues with a
finite drift due to their mutual repulsion,

\begin{eqnarray}
{\partial P\over\partial Y}   =
\sum_n {\partial \over \partial\mu_n}
\left[ {\partial \over \partial\mu_n} +
 \sum_{m\not=n}{\beta \over {\mu_m-\mu_n}} + \gamma \mu_n \right] P
\end{eqnarray}

($\beta=1,2$ for Hamiltonians with and without time-reversal, 
respectively). Here $\gamma$ is an arbitray parameter \cite{ps}.  
The evolution reaches a steady state when 
${\partial P/\partial Y}\rightarrow 0$; $P(\mu)$ in this limit is 
given by a WD distribution, 
$P(\mu) = \prod_{i< j}|\mu_i - \mu_j|^{\beta}
{\rm e}^{-{\gamma\over 2}\sum_k \mu_k^2}$.

The spectral correlations for MIT can now be obtained by using following 
analogy. 
The eq.(3) is same as the equation governing an evolution  
 of the distribution $P$ of the eigenvalues, of a Hamiltonian
$H=\sqrt{f} (H_0+\lambda V)$ with $\lambda^2={\rm e}^{2\gamma (Y-Y_0)}-1$ and 
$f={\rm e}^{-2\gamma (Y-Y_0)}$, from an initial state $H_0$ to  WD ensemble. 
The transition, referred as WDT  later on, is caused by  
a random perturbation $V$, taken from a WD ensemble (of variance $\gamma$),  
and of strength 
$\lambda$ \cite{dy,fh}; the elements $H_{kl}$ are thus Gaussian 
distributed with a variance $h_{kl}= (1-f)/2\gamma$,  
$h_{kk}=1/2\gamma$ and same mean for all of them. 
The transition (WDT) to
equilibrium, with $Y-Y_0$ as the evolution parameter, is rapid, discontinuous
for infinite dimensions of matrices \cite{dy}. 
But for small-$Y$ and large $N$, a smooth crossover can be seen in 
terms of a rescaled parameter $\Lambda= (Y-Y_0)/\Delta_{\eta}^2$, with 
$\Delta_{\eta}=\Delta N/\eta$ as the mean-level spacing in the correlated 
region of the spectrum at $Y-Y_0$, $\eta$ as the correlated "volume" and 
$\Delta\equiv \Delta(\mu,Y)$ as 
the mean level spacing of the whole spectrum. 
The intermediate states, corresponding to finite, non-zero $\Lambda$-values, 
during the crossover are known as Brownian ensembles (BE) \cite{dy,fh}  or 
Rosenzweig-Porter ensembles (RPE) \cite{shep,ks,fgm}. 
The similar evolution equations of $P$ for MIT and WDT, imply a similarity 
in their eigenvalues distributions for all $Y$-values and 
thereby correlations for all $\Lambda$-values, under similar  
initial conditions (that is, 
$P(\mu,Y_0)$ same for both the cases although $\rho(H,Y_0)$ may be 
different). 
In finite disordered systems, therefore, a continuous family of intermediate
statistics exists between metal and insulator which can be 
described by the BE for the corresponding $\Lambda$-value, occuring during 
Poisson $\rightarrow$ WD ensemble type WDT.

	 The behaviour of the mean level spacing $\Delta$ and the correlation  
volume $\eta$ divide the level-statistics for the large 
BE ($N \rightarrow \infty$)
into three regions \cite{shep}:  

(i) {\bf {$N^2 (Y-Y_0) \rightarrow 0$}}:  
 $\Delta \propto N^{-1}$ and $\eta<N$, giving  
 $\Lambda \rightarrow 0$ and Poisson 
statistics, 

(ii) {\bf $N^2 (Y-Y_0) \rightarrow \infty$}:   
$\Delta \propto N^{-1/2}$ and $\eta>N$ giving   
  $\Lambda \rightarrow \infty$ and  WD statistics.  

(iii){\bf $N^2 (Y-Y_0)=1/2c$, with $c$ as an arbitrary constant with respect to 
$N$}: although $\Delta$ still behaves as 
$o(1/N)$, but now  $\eta \approx N$ thus 
giving $\Lambda=1/2c\pi$ (referred as $\Lambda^*_{BE}$). 

The $\Lambda=\Lambda^*_{BE}$ therefore corresponds to a third statistics, 
intermediate between Poisson and WD ensemble     
 and is known as the critical Brownian ensemble (CBE). This being the 
case for arbitrary values of $c$, an infinite family of CBE occur during WDT. 
 The presence of such a family can 
be seen from any of the fluctuation measures for WDT. 
One traditionaly used meausure 
in this regard is relative behaviour of the tail of nearest-neighbour 
spacing distribution $P(s)$, defined as $\alpha (\delta,\Lambda) = 
\int_0^\delta (P(s)-P_w(s)){\rm d}s/\int_0^\delta (P_p(s)-P_w(s)){\rm d}s$ 
with $\delta$ as any one of the crossing points of $P_w(s)$ and $P_p(s)$ 
(here subscript $w$ and $p$ refer to the WD case and Poisson case respectively).  
In the limit $N\rightarrow \infty$, $\alpha=0$ and $1$ for WDE and 
Poisson limit respectively.  
The FIG.1 shows the numerically obtained behaviour of $\alpha$ with respect 
to $|c-c^*|$ (for $\delta\approx 2.02$), with 
$c^*$ corresponding to one of the CBEs. The convergence of all the points on 
two branches    
for different $N$-values confirms the existence of a CBE at $c^*$ with a 
critical exponent $\nu \rightarrow \infty$ (as $\alpha(c) \approx\alpha(c^*)+  
constant.|c-c^*|N^{1/\nu}$ near $c^*$). The $N$-independence of $\alpha (c)$ 
 also indicates the possibility of infinitely many $c^*$ (and its arbitrarity) 
and 
therefore 
an infinite family of CBEs.              

 For disordered systems with infinite system size $L$, 
the states are critical near a particular disorder $W^*$ (or energy), termed as 
critical point, with a correlation length $\zeta(W) \propto |W-W^*|^{-\nu}$, 
$\nu$ as the critical exponent. At the critical point, $\zeta \approx L$ and 
the critical value of the parameter $\Lambda$ (referred as $\Lambda^*_{AH}$) 
depends on the 
dimensionality; $\Lambda^*_{AH}=(Y-Y_0)/\Delta_{\eta}^2$, with both $Y$ and 
$\Delta_{\eta}$  
dimensionality-dependent. A knowledge of $\Lambda^*_{AH}$ can then be used to map 
the critical level statistics at MIT for various 
dimensions $d >2 \rightarrow \infty$ to the infinite 
family of critical Brownian ensembles (CBE).

	The identification of the appropriate CBE corresponding to the 
 critical Anderson Hamiltonian (CAH) requires $\Lambda^*_{BE}=\Lambda^*_{AH}$, 
and, thus  a prior knowledge of  
$\Lambda$-value associated with CAH. 
We consider here one example (this case is also used in 
our numerical analysis). Consider an Anderson system with 
the Gaussian disorder, same for each site, and random or non-random hopping
between nearest neighbours.       
The corresponding ensemble measure can be described by eq.(1) with  
 $h_{kk}=W^2/2$, $b_{kk}=0$ and $h_{kl}=W_1^2/2$, $b_{kl}=t$ for $\{k,l\}$
pairs representing hopping,  $h_{kl}\rightarrow 0$ and $b_{kl}
\rightarrow 0$ for all $\{k,l\}$ values corresponding to dissconnected sites.  
This gives, by using eq.(2),  
$Y=-(N/2 M\gamma){\rm ln}\left[|1-\gamma W^2| |1-2\gamma W_1^2|^{\beta z/2}  
|t+\delta_{t0}|^{\beta z} \right] + C$. 
 Here $M=\beta N(N+z \delta_{t0}+2-\beta)/2$ with $z N$ as the  
number of the connected sites 
(nearest-neighbours)  which depends on the topology and the dimensionality 
$d$ of the system. Analogous to WDT, a 
rescaling of $Y-Y_0$ would give the parameter for smooth transition:
$\Lambda= 2(a-a_0)\beta^{-1}{(\zeta/L)}^d$ with 
$a(\alpha,t) \equiv 
{\rm ln}\left[|1-\gamma W^2| |1-2\gamma W_1^2|^{\beta z/2} 
|t+\delta_{t0}|^{\beta z} \right]$,  
and $a_0$ as the value of $a$ at $Y_0$. 
Here $\Delta$ behaves as $O(N^{-1/2})$
 (as $N^2(Y-Y_0) \rightarrow \infty$)
 and $\eta={\zeta}^d$ 
with $\zeta$ as the localization length which depends 
 on disorder $W$ and can be determined by using wavefunction statistics 
(e.g. inverse participation ratio).  
As disorder decreases, the $\zeta(W)$ increases resulting 
in a smooth increase in $\Lambda$ in finite systems, and, thereby the statistics 
intermediate between Poisson and WDE. However for infinite system sizes,   
 $\Lambda \rightarrow 0$ and the statistics is Poisson as long as $\zeta<L$. 
 But at a certain disorder strength, $\zeta \sim L$, which makes 
$\Lambda$ size-independent thus giving its critical value $\Lambda^*_{AH}  
\approx 2(a-a_0)\beta^{-1}$. 
The level statistics of the system at 
this disorder strength is given by the CBE with the same   
$\Lambda$-value and is critical due to (i) being different from both 
Poisson and WD behaviour, (ii) its invariance as 
$N\rightarrow \infty$. 
 Now for a further decrease in disorder, $\zeta>L$, 
as a result       
 $\Lambda \rightarrow \infty$, and the system has a WD statistics. 
However, for an infinitely long one dimensional lattice with $z$ nearest 
neghbours, 
the statistics always remains Poission irrespective of disorder 
strength. This is because   
$\zeta\approx z^2$ giving $\Lambda \approx z^2/L 
 \rightarrow 0$ for $L \rightarrow \infty$ unless there is a 
very long range connectivity in the lattice (i.e $z\approx \sqrt{L}$).  
However 
 a crossover from Poisson 
to WD ensemble can be seen for finite $L$ by varying the ratio $z^2/L$.

In general, $\Lambda$ will be a function of coordination number,  
disorder strength, hopping rate and  dimensionality of the lattice as well as 
the level-density of the system.  
The boundary conditions/ topologies, leading to different sparcity and 
coordination 
numbers, may therefore result in different critical level statistics  
even if degree of disorder, hopping rate and dimensionality is same; this is
in agreement with numerical observations \cite{bran} and analytical study for
2D systems \cite{ky}. Similarly different dimensions, 
can lead to diffenent local level-densities and thereby different critical points 
$\Lambda^*$.

Many results for the spectral fluctuations of the WDT with Poisson 
ensemble as an initial state are already known \cite{ap} and can 
directly be used for the corresponding measures for the MIT in 
different disordered systems.  
For example, consider the 2-level density correlator $R_2(r;\Lambda)$;  
$R_2(r;\Lambda)= <\nu (\mu_1,\Lambda)\nu(\mu_2,\Lambda)>/<\nu>^2 
 = { N! \over {(N-2)!}}\int P(\mu, \Lambda){\rm d}\mu_{3}..{\rm d}\mu_N$. 
Here $\nu(\mu,\Lambda) = N^{-1} \sum_i \delta (\mu-\mu_i)$ is the density of
eigenvalues  and $<..>$ implying the ensemble average. 
The $R_2$ for the Anderson transition in presence of a magnetic field, 
can therefore be 
given by the $R_2$ for WDT between Poisson $\rightarrow$ GUE ensemble 
\cite{ap,ks}; $R_2= 1 - Y_2$ where  
\begin{eqnarray}
Y_2(r;\Lambda)=-{4 \Lambda\over r} 
\int_0^\infty {\rm d}u \; F \; {\rm e}^{-2\Lambda u^2-4\pi\Lambda u}  
\end{eqnarray} 
with $F={\rm sin}(ur) f_1 - {\rm cos}(ur) f_2$,  
$f_1=(2/z)[I_1(z)-\sqrt{8u/\pi}I_2(z)]$,
$f_2=(1/u)[I_2(z)-\sqrt{2u/\pi}I_3(z)]$,  
 $z=\sqrt{32\pi \Lambda^2 u^3}$ and $I_n$ as the $n^{\rm th}$
Bessel function.
Here $R_2(r,\infty)= 1- {({\rm sin}^2(\pi r)/\pi^2 r^2)}$
 and $R_2(r,0)=1$ corresponding to metal and insulator regime
respectively. A substitution of $\Lambda=\Lambda^*$ in eq.(4)
	will thus give the $R_2$  for the CAH.  
Similarly the nearest-neighbour spacing distribution $P(s)$ for the MIT 
can be given by using the one for the BE  
during Poisson $\rightarrow$ GUE transition \cite{to}:
$P(s;\Lambda) \propto (2\pi\Lambda)^{-1/2} s {\rm e}^{-s^2/8\Lambda}
\int_0^\infty {\rm d}x\; x^{-1} {\rm e}^{-x-x^2/8\Lambda} 
{\rm sinh}(xs/4\Lambda)$. 
Further the good agreement of the numerically obtained 
$P(s)$ for the CAH without time-reversal symmetry    
with that of a CBE (with same $\Lambda$-value) 
reconfirms our analytical result; the FIG.2(a,b) show the behaviour  
of the CAH in presence of a random hopping    
(CAH-I with $W=8.15$, $W_1=1$, $t=0$, and, thereby $\Lambda=10.6$) and a  
non-Random hopping (CAH-II with $W=21.3$, $W_1=0$, $t=1$, thus 
$\Lambda \approx 0.53$), respectively along with their CBE analogs ($c$ 
determined by 
the relation $\Lambda^*_{AH}= 
 \Lambda^*_{BE}=(2c\pi)^{-1}$); 
see \cite{xx} for the details on the numerics.

An important characterstic
of  critical level statistics  is the level compressibility $\chi$, 
 $\chi(\Lambda) \approx
1-\int_{-\infty}^{\infty} Y_2 (r;\Lambda) {\rm d}r$; 
 $\chi=0,1$ in the metallic  and the insulator phase, respectively,  
and takes an intermediate value at the hybrid phase near the critical point.
The eq.(4) can  be used to obtain
$\chi(\Lambda)= 1-4 \pi \Lambda\int_0^\infty {\rm d}u
f_1 (z){\rm exp}[- 2\Lambda u^2 - 4\pi\Lambda u]$.
The critical region, with its finite $\Lambda$ value ($=\Lambda^*$), 
will thus have
a  level compressibility different from 
both metal and insulator regimes; a $0 < \chi <1$ value is supposed to be 
an indicator of the multifractal nature
of the eigenvectors.

The  compressibility of the spectrum can also be
seen from the "number variance" $\Sigma_2$
which describes the variance in the number of levels in an interval of $r$ 
mean level spacings. The $\Sigma_2(r)$, basically a measure of the "spectral 
rigidity" is related to the compressibility: 
${\rm lim}\;  r\rightarrow \infty \; \Sigma_2 \approx \chi r$. 
 The FIG.3 shows the 
numerical behaviour for the $\Sigma^2(r;\Lambda)$
 for the CAH-I and CAH-II along with their CBE analogs; the good 
agreement in each case re-verifies our claim about the similarity 
between the MIT and WDT.  
Furthermore FIG.3. also shows the fractional behaviour of $\chi$ for CBE,   
which is in agreement with the $\chi$-form given in the preceeding paragraph.

 The statistical measures for the Anderson transition in
presence of a time-reversal symmetry can similarly be obtained by using their
 equivalence  to WDT from Poisson $\rightarrow$ GOE ensemble. 
However due to the technical difficulties \cite{ap}), 
only some approximate results are known for the latter case.
for example, the $R_2$ for small-$r$
can be given as $R_2(r,\Lambda) \approx (2r-1)^{1/2}
J_{1/3}\left((2r-1)^{3/2}/3\Lambda \right) {\rm e}^{r/2\Lambda}$
with $J(z)$ as the Bessel function (obtained by solving eq.(17) of \cite{ap}). 
Similarly for large-$r$ behaviour,
$R_2$ can be shown to satisfy the relation (see eq.(21) of \cite{ap})
$R_2(r,\Lambda)= R_2(r,\infty) + 2\beta \Lambda
\int_{-\infty}^{\infty} {\rm d}s {R_2(r-s;0)-R_2(r-s;\infty)/ 
(s^2 + 4\pi^2 \beta^2 \Lambda^2 )}$.
By taking $\beta=1$ and $R_2(r,\infty)=
 1- {{\rm sin}^2(\pi r)/ \pi^2 r^2} -
\left(\int_r^\infty {\rm d}x {{\rm sin}\pi x / \pi x}\right)
\left({{\rm d}\over {\rm d}r}{{\rm sin}\pi r / \pi r}\right)$
(GOE limit), the $R_2(r,\Lambda)$ for the transition in presence of a
TRS can be obtained; $R_2(r,\Lambda) \approx R_2(r,\infty) +
{4\Lambda /( r^2 + 4\pi^2 \beta^2 \Lambda^2)}$ .
The lack of the knowledge of $R_2(r,\Lambda)$ for entire energy-range 
 handicaps us in providing an exact form of $\chi$.  
However the $P(s)$ for this case can be given by using the one for a  
 BE (RPE) during Poisson $\rightarrow$GOE transition \cite{to}:
$P(s,\Lambda)=(\pi/8\Lambda)^{1/2} s {\rm e}^{-s^2/16 \Lambda}
I_0(s^2/16\Lambda)$, with $I_0$ as the Bessel function.

	In the end, a comparison of our results with some of the 
past studies is crucial.
The presence of a fractional  
compressibility in an ensemble essentially same as CBE and its possibility 
as a model for CAH was also suggested 
in \cite{hp} which was 
later on contradicted in \cite{fgm}.
  The claim in \cite{fgm} is disproved by our numerical 
studies of both the transitions, showing similar behaviour for 
compressibility, besides other fluctuation measures at the critical point. 
The numerical work therefore reaffirms our claim  based on the 
exact analytical 
work: 
(i) the level-statistics for MIT in disordered systems with a 
Gaussian disorder can 
be described by the same for WDT with a Poisson initial condition,  
(ii) the level-statistics in the disordered systems is indeed governed by a  
single scaling parameter.

\end{multicols}

\section{Figure Caption}

\section*{Captions}
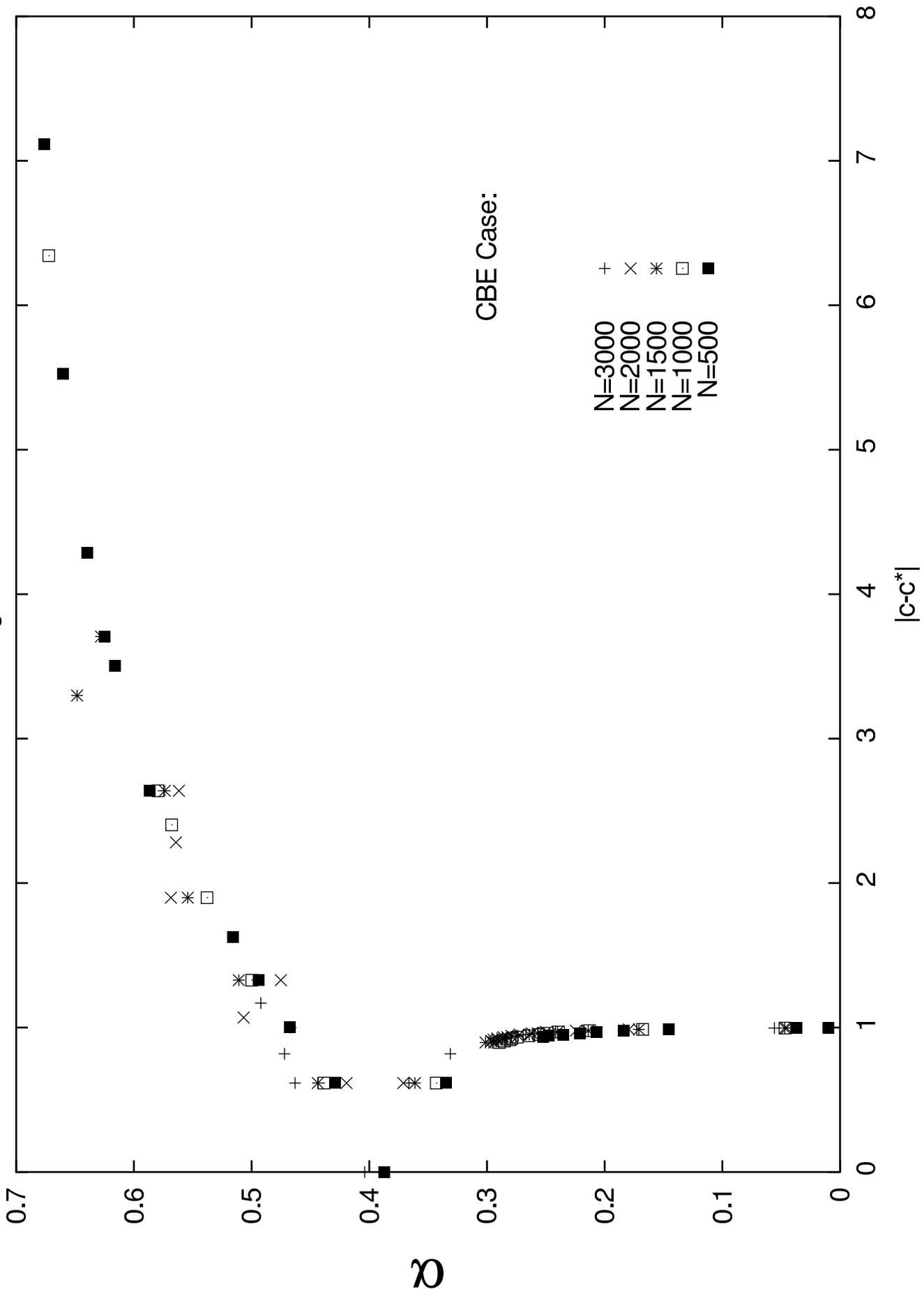
\noindent Fig. 1. $\alpha$ vs $|c-c^*|$ for WDT
\\ \\
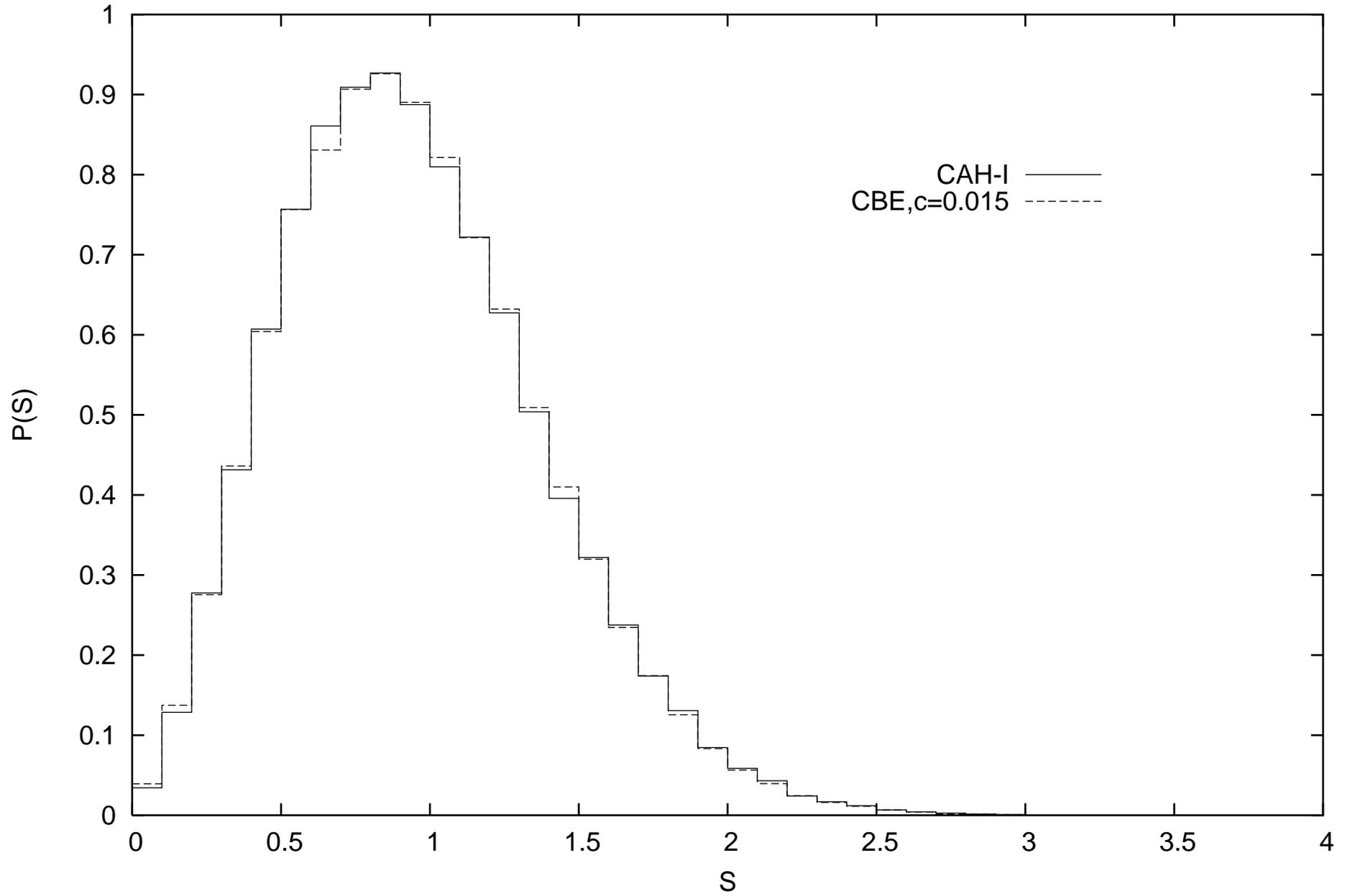
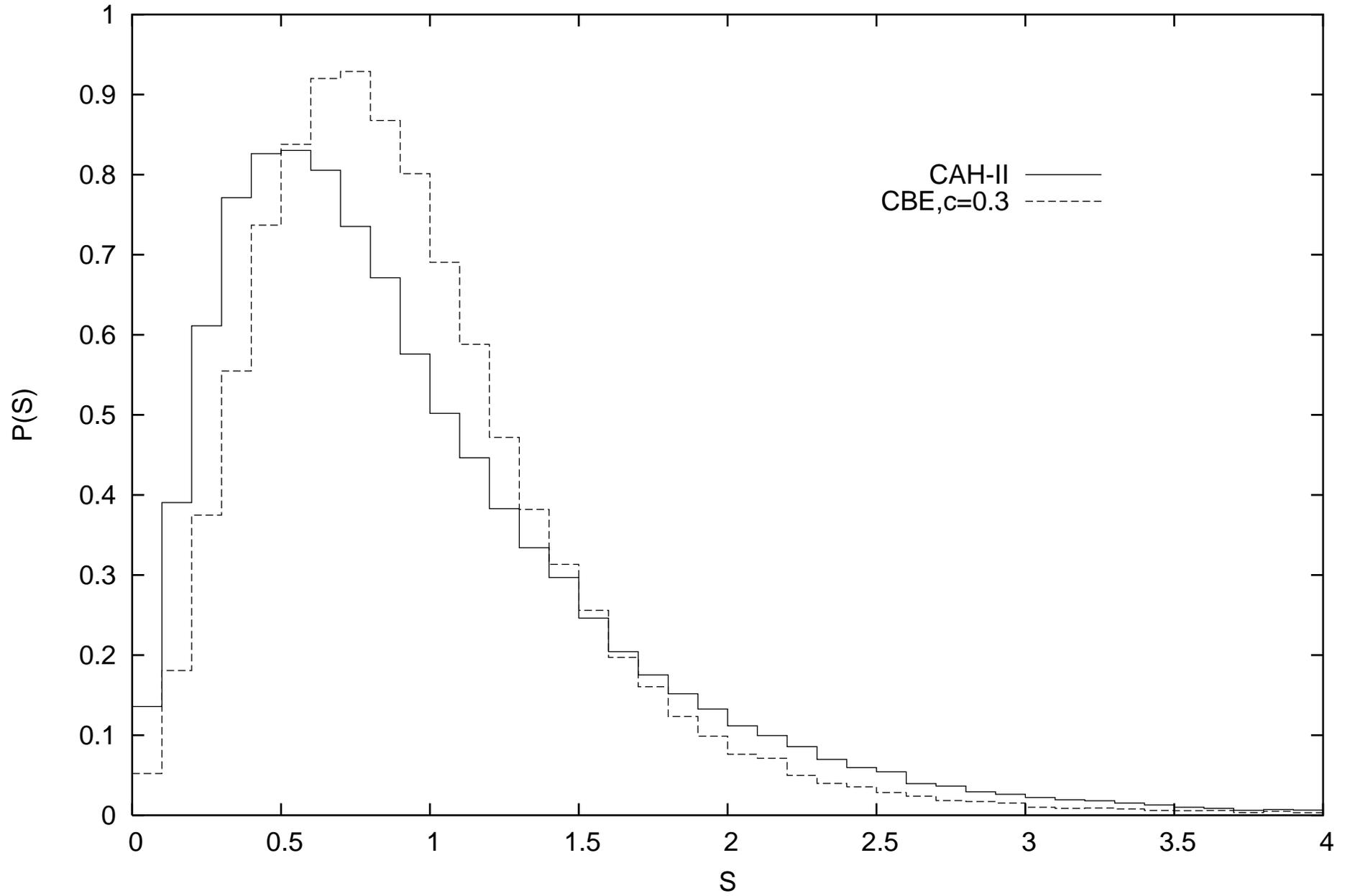
\noindent Fig. 2.  $P(S)$ vs $S$ for 
(A) CAH-I and the corresponding CBE ($c=0.015$),
(b) CAH-II and the CBE with $c=0.3$. 
\\ \\ 
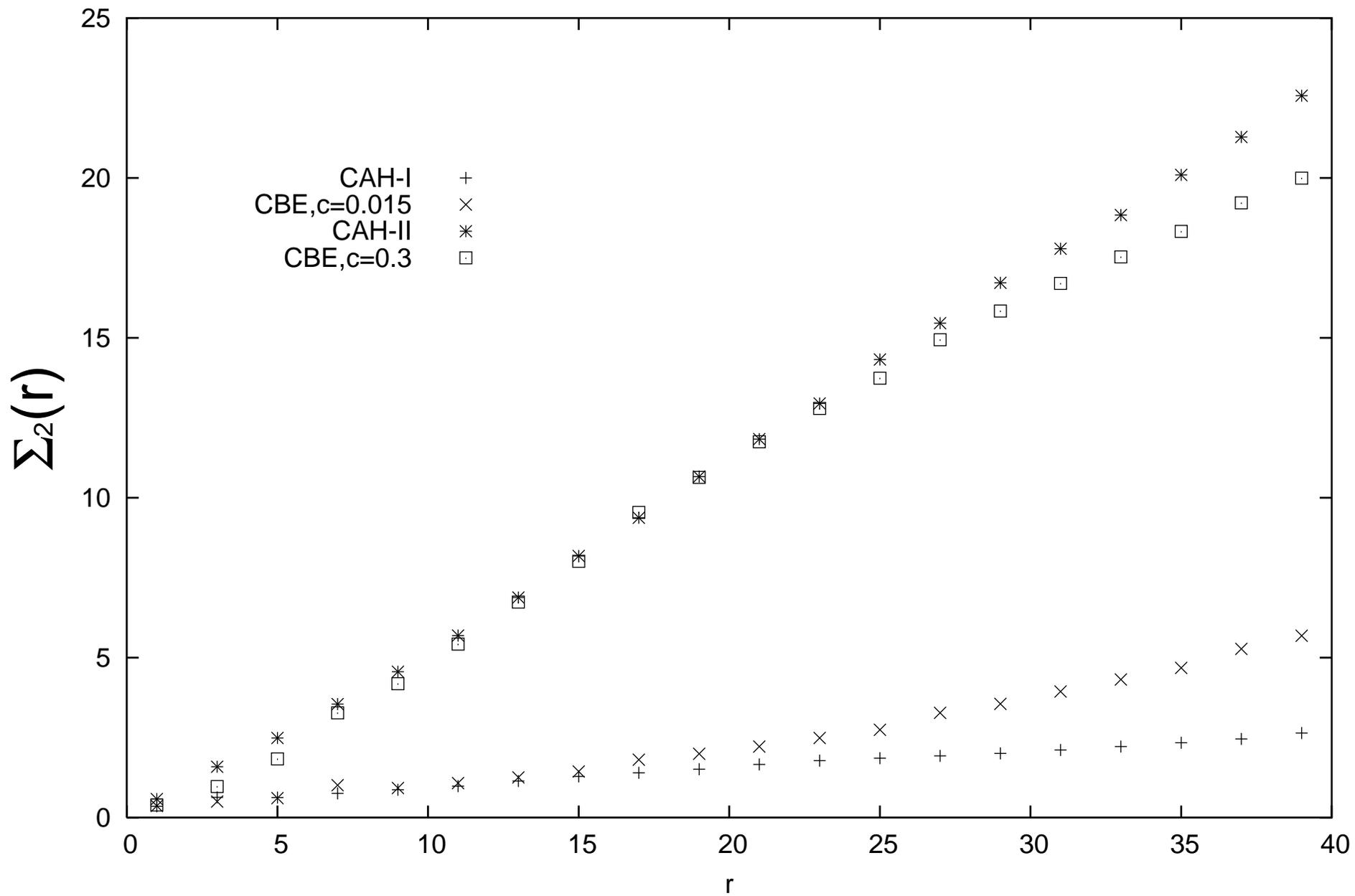
\noindent Fig. 3.  $\Sigma^2$ vs $r$ for 
(i) CAH-I and  CBE with $c=0.3$, 
(ii) CAH-II and the CBE with $c=0.015$.


\begin{references}
\bibitem{xx}
The ensemble of matrices considered in each BE ($N=3000$) and 
CAH ($L=13$) case consisted of 
$300$ matrices.  
To avoid mixing of the levels with 
different transition rates,     
we used only $27 \%$ of 
the levels from middle of the spectrum of each matrix.
\end{references}
\end{document}